# Field-induced low-temperature electronic specific heat of boron nitride nanotubes

Feng-Lin Shyu

Department of Physics, R.O.C. Military Academy, Kaohsiung 830, Taiwan, R.O.C.

November 26, 2015

E-mail: fl.shyu@msa.hinet.net, Phone: +88677425024, Fax: +88677194170

**Abstract**

We use the tight-binding model to study the effect of transverse electric field on the low-temperature electronic specific heat ($C_v$) for armchair and zigzag boron nitride nanotubes (ABNNTs and ZBNNTs). For wide-band-gap BNNTs, electric field could significantly modulate their energy dispersions and shift many electronic states close to the Fermi energy. Under a critical electric field ($F_c$), the density of states show special peak structures and the vanishing specific heat at zero field jumps to a giant one. $C_v$, at $F_c$'s, has a value comparable to that of the phonon specific heat and reveals strongly non-linear dependence on temperature. The critical field strength and the value of giant specific heat are closely related to nanotube's geometry. In the presence of $F_c$'s, the extra longitudinal magnetic flux could enhance the value of $C_v$ again at low temperature for ZBNNTs, whereas it is not always true for ABNNTs.

## 1. Introduction

Carbon nanotubes (CNTs) described as a rolled-up graphene sheet have stirred many intensive studies owing to their unique physical properties and widely potential application in nanodevices. Due to similar hexagonal symmetry in layered boron nitride, the existence of boron nitride nanotubes (BNNTs) was first theoretically predicted by Rubio *et al.* [1]. Following the theoretical prediction, synthesis in experiment was performed by Chopra *et al.* using an arc-discharge method [2]. Unlike the strongly geometry-dependent electronic structures of CNTs, semiconducting BNNTs have a wide-band-gap ranging from 4 *eV* to 5 *eV* independent of their geometry [3]. Based on band-gap engineering, these results anticipate more practical applications in nanoscaled electronic and photonic devices for BNNTs than CNTs.

Some previous studies showed that applying transverse electric field could modulate electronic structures of CNTs and BNNTs [4-6], except for mechanical deformation and impurity doping. As a result, band gap was reduced and even semiconductor-metal transition occurred. The giant Stark effect was further validated experimentally by Ishigami *et al.* through the bias-dependent scanning tunneling microscopy [7]. From the point of view of perturbation, the Stark effect could cause a strong coupling for neighboring subbands with nearly equivalent state energy. That would lead to a modulation of energy dispersions and further change optical and magnetic properties [8-11]. Since BNNTs are III-V compounds, their boron-nitride bonding with strong ionicity is more sensitive to the giant Stark effect than the C-C bonding in CNTs. Therefore, a significant change in electronic and physical properties induced by electric field is expected for BNNTs.

With the rapid advance of nano-devices, heat removal has become a crucial issue due to increased levels of dissipated power. Searching for novel materials that conduct heat well has become essential for designing the new integrated circuits and photoelectronic devices. Therefore, thermal properties such as specific heat and thermal conductivity have been extensively investigated in the past decades. The ability, for materials, to conduct heat is related to their atomic structures. Especially, rich changes in thermal properties of materials are revealed when they are structured on nanometer scale. The specific heat of nanomaterials coming from phonons ($C_{ph}$) or electrons ($C_v$) is determined by the details of phonon or electron spectra. The low-temperature behavior of the specific heat includes informations involving thermal excitations and the low-dimensional quantum confinement. A two-dimensional graphene, for instance, the ratio of $C_{ph}$ to $C_v$ at the low-temperature is about $10^4$, whereas it is down to $10^2$ for an one-dimensional CNT [12]. The specific heat, in these two carbon-related system, is primarily dominated by phonon's contribution even for $T \rightarrow 0$. It could be roughly explained by that the Fermi velocity in electron spectra is two orders of magnitudes greater than the sound velocity in phonon spectra. In order to increase the low-temperature electronic specific heat, reducing the Fermi velocity of electron spectra, i.e., increasing band-edge states near the Fermi energy, is required. Our previous study showed that applied transverse electric field, for BNNTs, could effectively change electronic energy dispersions around the Fermi energy. Critical electric field could enhance the magnitude of persistent current and change the magnetism [11]. It is predicted that electric field could also effectively modulate the specific heat of wide-gap BNNTs.

For BN-related systems, thermal properties have been studied for hexagonal boron nitride [13] and multi-walled BNNTs [14-15] in experiments. Specific heat of hexagonal BN at low temperature ($T < 10\ K$) showed a $T^3$-dependence while it revealed a $T^{2.4}$-dependence for thermal conductivity. The different $T$-dependence between specific heat and thermal conductivity was due to electron-phonon scattering. As for multi-walled BN-NTs, the $T$-dependence of thermal conductivity in experimental measurements reflected low-dimensional quantum confinement and revealed that BNNTs might be better thermal conductors than CNTs. For single-walled BNNTs, thermal properties have been studied theoretically [16-18]. The calculated results predicted that a higher specific heat for BN-NTs than that for CNTs. Moreover, thermal conductance of BNNTs exhibited a universal quantization at low temperature, independent of nanotube's geometry. As mentioned in the above, the low-temperature thermal properties, due to the wide-band-gap of BNNTs, are all dominated by phonon effects, thoroughly independent of electronic states. Our study thus intends to raise the electronic contribution to specific heat by applying external fields. It is expected that external fields could cause an increase of electronic states close to the Fermi energy that might largely enhance the contribution of electrons to low-temperature thermal properties.

In this work, we use $2p_z$ orbital tight-binding model to study electronic structures and low-temperature electronic specific heat for ABNNTs and ZBNNTs in the transverse electric field. Our study shows electric field strongly modulates energy dispersions and largely reduces band gap to a value comparable to the low-temperature thermal energy. The DOS reveals special peak structures causing a giant electronic specific heat, while electric field is at a critical strength. Furthermore, the extra magnetic field would re-enhance the giant specific heat. Modulations of external fields on electronic specific heat such as the magnitude and temperature-dependence are strongly dependent on the geometric structure of BNNTs.

## 2. The tight-binding model for electronic structures in external fields

A boron nitride nanotube is made of a monolayer hexagonal boron nitride (hBN) rolled up into a hollow cylinder. Its geometric structure could be built from a hBN and character-ized by two vectors, i.e., the first $\mathbf{R}_x = m\mathbf{a}_1 + n\mathbf{a}_2$ in the circumferential direction and the second $\mathbf{R}_y = p\mathbf{a}_1 + q\mathbf{a}_2$ along the longitudinal direction. Where $(m, n, p, q)$ are integers, $\mathbf{a}_1$ and $\mathbf{a}_2$ are primitive lattice vectors of a hBN. Due to the orthogonality $\mathbf{R}_x \cdot \mathbf{R}_y = 0$, the parameters $(m,n)$ uniquely define the geometric structure of a BNNT. The radius and chiral angle are $r = R_x/2\pi = b\sqrt{m^2 + mn + n^2}/2\pi$ and $\theta = tan^{-1}[-\sqrt{3}n/(2m+n)]$, respectively.

$b = 1.45$ Å is B-N bond length. $(m,m)$ armchair boron nitride nanotubes (ABNNTs) and $(m,0)$ zigzag boron nitride nanotubes (ZBNNTs) belong to achiral systems. A $(m,m)$ ABNNT has $r = 3mb/2\pi$ and $\theta = -30°$. They are $r = \sqrt{3}mb/2\pi$ and $\theta = 0°$ for a $(m,0)$ ZBNNT. The number of atoms in a unit cell is $N_u = 4\sqrt{(p^2 + pq + q^2)(m^2 + mn + n^2)/3}$, and both the ABNNT and ZBNNT have the same $N_u$ ($=4m$).

The hermitian Hamiltonian matrix, in the tight-binding model, is built from the subspace spanned by the $N_u$ wave functions of $2p_z$ orbitals. In the presence of electric and magnetic fields, the Hamiltonian including nearest-neighbor interactions is given by

$$H = \sum_i \epsilon_i c_i^+ c_i + t_0 \sum_{i,j} e^{i(2\pi/\phi_0)\int_i^j \mathbf{A}\cdot\mathbf{dr}} c_i^+ c_j, \qquad (1)$$

where $\epsilon_i = E_i + Fr cos\alpha_i$ is the on-site energy due to the $2p_z$ atomic orbital and the external electric field. $\alpha_i$ is the angle between the position vector of the $i$th atom and the transverse electric field $\mathbf{F}$ (unit $eV/$Å). $t_0 = -2.92$ $eV$ is the nearest-neighbor hopping integral. The $2p_z$ on-site energies ($E_i$) of boron atom and nitrogen atom are $E_B = 4.78$ $eV$ and $E_N = 0.48$ $eV$, respectively. $c_i^+$ ($c_i$) is the creation (annihilation) operator. $exp[i(2\pi/\phi_0)\int_i^j \mathbf{A}\cdot\mathbf{dr}]$ is the magnetic phase, where $\mathbf{A}$ is the vector potential. When the transverse electric field and the longitudinal magnetic field ($\mathbf{B}$) are applied, the period along the nanotube's axis is not destroyed. The longitudinal wave vector ($k_y$) is still a good quantum number and specify energy dispersions of BNNTs. The first Brillouin zone has the range $-\pi/R_y \leq k_y \leq \pi/R_y$. For ABNNTs and ZBNNTs, the band structures could be calculated by diagonalizing the $4m \times 4m$ hermitian Hamiltonian matrix.

All BNNTs without external field are wide-gap semiconductors but the energy dispersions are strongly dependent on geometric structures. Here we give a brief review for band structures of BNNTs [11]. For the ABNNTs, conduction and valence bands are symmetric about the Fermi energy ($E_F = 2.63$ $eV$). Most of subbands are doubly degenerate except the first (closest to $E_F$) and the last subbands. The band-edge states are located at $k_y = 2/3$ (unit $\pi/R_y$). Energy spacing decreases with increasing wave vector $k_y$ and all subbands are merged together at $k_y = 1$. As for the ZBNNTs, energy dispersions show similar band-symmetry and degeneracy, but the two singlet states correspond to the fifth and last subbands. In contrast to an ABNNT, the different features in band structures of a ZBNNT include (I) the band-edge states are at $k_y = 0$ and (II) all subbands, at $k_y = 1$, are merged into one doubly and four four-fold degenerate states.

While a BNNT (or CNT) is threaded by uniform magnetic field along the nanotube axis, the induced magnetic phase could change the band degeneracy, energy dispersion, energy spacing, and band gap. With increasing magnetic flux, electronic structures at $\phi = \phi_0$ are restored to those at $\phi = 0$, i.e., Ahanorov-Bohm (AB) oscillation. Compared with the sharp sensitivity of energy dispersions in CNTs to magnetic field, magnetic field just slightly changes band-width and band gap for BNNTs. According to our previous studies [12], the maximum change in the magnitude of band gap is about 0.08 $eV$ which is much smaller than band gap ($\sim 4.5\ eV$) at $\phi = 0$. Moreover, magneto-energy gap weakly depends on the nanotube's radius. Thus, effective modulations of energy dispersions should be considered by other methods.

BNNTs, unlike CNTs, are III-V compound nanotubes with strong ionicity, their electronic structures are expected to be much more sensitive to applied electric field than magnetic field. While the transverse electric field is applied, the coordinate-dependent electric potential breaks the rotational symmetry such that the transverse momenta are no longer good quantum numbers. The neighboring subbands with nearly equivalent energy would be coupled and the coupling is getting stronger with increasing field strength that leads to a strongly $k_y$-dependent energy dispersion. The magnitude of electric field leading to band gap with a value smaller than $10^{-3}\ eV$ is considered, since the low-temperature ($T < 5$ $K$) thermal properties are our concerns. Electric field not only reduces band gaps but also significantly changes energy dispersions (or density of states). For a (12,12) BNNT, Fig. 1(a) shows the variation of energy dispersion with various $F$'s which could reduce band gap to $10^{-3}$ $eV$. Energy dispersion is changed from a parabolic into a two-local-minima structure except the reduction of band gap while electric field increases from $F = 0.3078$ to $F = 0.3081$. There is a tiny energy difference between the two local minima that is reflected in the DOS (Fig. 1(b)). Moreover, the symmetry of conduction and valence subbands about the Fermi energy is unaltered.

The characteristics of field-induced energy dispersions are unveiled in the DOS which could further give a detailed elucidation to thermal properties. The special (divergence) structures of the DOS imply a drastic change in thermal properties; the magnitude and $T$-dependence of the low-temperature electronic specific heat are determined by peak positions and heights of the DOS. It is defined as

$$D(\omega) = \frac{1}{N_u} \sum_{h,\sigma} \int_{1stBZ} \frac{dk_y}{2\pi} \frac{1}{\pi} \frac{\delta}{[\omega - E^h(F, k_y; \sigma, \phi)]^2 + \delta^2}, \qquad (2)$$

where $h = c, v$ represents conduction or valence bands and the phenomenological broadening parameter is $\delta = 10^{-4}$ $eV$. For a (12,12) BNNT, solid black curves ($F = 0.3078$) in Fig. 1(b) show that the special structure for $E > E_F$ diverges in the form $1/\sqrt{E^c(k_{ye}) - \omega}$ from the concave upward edge-state with a wave vector $k_{ye} = 0.688$. Oppositely, the other ($E < E_F$) is in the form $1/\sqrt{\omega - E^v(k_{ye})}$ from the concave downward edge-state. While critical field $F_c^A = 0.3079$ is applied, the parabolic subbands are flattened leading to much sharper peak structures (solid red curves); that would make a big contribution to the low-temperature thermal properties. As electric field continuously increases from $F = 0.3080$ to $F = 0.3081$, the two-local-minima states are created showing two lower split peak structures in the DOS. The number of low-temperature thermal electronic states is also reduced.

While the radius of ABNNTs increases, the features of energy dispersions are almost unchanged. But the stronger electric potential makes electronic states closer to the Fermi energy, and then the band gap and the critical electric field are thus effectively reduced, as shown in Figs. 1(c) and 1(e) for (13,13) and (14,14) BNNTs. Moreover, the critical field is no more determined by the peak height of the DOS; the peak position is also a key factor. For example, critical field of a (13,13) BNNT is $F_c^A = 0.2806$ because its edge-states are closer to $E_F$ than those of $F = 0.2805$; that could provide much more low-temperature electronic states.

As for $(m,0)$ ZBNNTs, (21,0), (22,0), and (23,0) BNNTs having the same radii as the mentioned ABNNTs are chosen as a comparison. The $F$-dependent electronic structures have strong geometric dependence described as follows. The doubly degenerate states closest $E_F$ are split into two singlet states by electric field, as shown in Fig. 2(a) for a (21,0) BNNT. As electric field gradually increases, energy dispersion oscillates and shows stronger $k_y$-dependence. Moreover, edge-states are shift from $k_y = 0$ to larger $k_y$'s and band gap is reduced as shown in the DOS (Fig. 2(b)). Each pair of peaks in the DOS is from two split singlet states while it is from two local minima of a singlet state for ABNNTs. Similarly, there also exists a critical electric field ($F_c^z = 0.3040$) inducing a sharp peak structure in the DOS and making a contribution to low-temperature thermal properties. It is noticed that for ABNNTs, the reduction of band gap by electric field is enhanced with increasing their radii. However, the rule is not true for ZBNNTs shown in the DOS from Figs. 2(b) and 2(d). For a $(m,0)$ ZBNNT, a nanotube with $m = 3I$ ($I$ an integer), e.g., a (21,0) BNNT, has the smallest band gap while one with $m = 3I + 1$ (a (22,0) BNNT) shows the largest band gap. The above rule also holds even for larger $m$, e.g., (24,0), (25,0), and (26,0) BNNTs. Therefore, $F$-dependent band gaps, for ZBNNTs, are profoundly related to geometric structures that would be fully reflected in the low-temperature electronic specific heat.

## 3. Field-modulated electronic specific heat

The electronic specific heat is defined as the derivative of the total energy with respect to temperature. In the presence of electric and magnetic fields, the total energy per mole carrier at temperature $T$ is written as

$$U(T) = \frac{3\sqrt{3}b^2 N_A}{8\pi r} \sum_{\sigma,h} \int_{1stBz} \frac{dk_y}{2\pi} [E^h(F, k_y; \sigma, \phi) - \mu] f(E^h(F, k_y; \sigma, \phi), T), \quad (3)$$

where $N_A$ is the Avogadro's number, and $f = 1/\{exp[\beta(E^h(F, k_y; \sigma, \phi) - \mu)] + 1\}$ is the Fermi-Dirac distribution function with $\beta = 1/k_B T$ and $k_B$ denotes the Boltzmann constant. The electronic specific heat at temperature $T$ could be calculated by

$$\begin{aligned} C_v(T) &= \frac{\partial U(T)}{\partial T} \\ &= \frac{3\sqrt{3}b^2}{8\pi r} \frac{N_A}{k_B T^2} \sum_{\sigma,h} \int_{1stBz} \frac{dk_y}{2\pi} \frac{(E^h(F, k_y; \sigma, \phi) - \mu)^2 e^{\beta(E^h(F, k_y; \sigma, \phi) - \mu)}}{(1 + e^{\beta(E^h(F, k_y; \sigma, \phi) - \mu)})^2}. \quad (4) \end{aligned}$$

In the above equation, the term related to the derivative of chemical potential with respect to temperature is omitted, since the chemical potential $\mu$, due to the $\pi$-band symmetry, is fixed at $E_F$ for arbitrary field strength and temperature.

The electronic specific heat, due to the wide band gap, vanishes in the absence of electric field. Our main study thus is how electric field modulates the electronic specific heat at low temperature ($T \leq 5\ K$). For a (12,12) BNNT, the detailed $F$-dependent specific heat at $T = 1\ K$ is shown in the inset of Fig. 3(a). The specific heat keeps vanishing for insufficient field strength. While the first critical electric field is applied ($F_{1c}^A = 0.3079$), energy dispersions are flattened and edge-states are closer to the Fermi energy. Band gap is comparable to the thermal energy ($\sim 10^{-4}\ eV$) at $T = 1\ K$. There are sufficient electronic states to initiate low-temperature specific heat with a jump structure which has a value $C_v = 153$ (unit $\mu J/mole\ K$, here and henceforth). Here, $C_v$, due to critical electric fields, jumps from a zero (or very tiny) value to a very large one which is called the giant electronic specific heat (GESH). With continuously increasing electric field, the edge-states become two-local-minima states with higher curvature that reduces the DOS and thus the value of $C_v$. As temperature gradually increases, more and more thermally excited electrons occupy the electronic states above the chemical potential, and leaving holes at original states. These two kinds of excited carriers equivalently make contribution to the specific heat. Therefore, the contribution to the specific heat made by the Fermi function is extended to wider energy range. The jump structure in the $F$-dependent specific heat reveals higher peak and wider width with increasing $T$.

Experimentally, the field-enhanced specific heat is more easily performed for larger BNNTs owing to the smaller applied electric field. Thus, the $F$-dependent specific heat with varying nanotube's radius is further studied. We first choose $(m,m)$ ABNNTs with $m$=12, 13, and 14 as a model study. The comparison among Figs. 3(a), 3(b) and 3(c) shows that the critical field initiating the first jump structure in $C_v$ decreases with increasing nanotube's radius, e.g., $F_{1c}^A = 0.2806$ for a (13,13) BNNT and $F_{1c}^A = 0.2576$ for a (14,14) BNNT, respectively. Besides, $C_v$-$F$ curves of a larger BNNT also reveal richer special structures within strong field region, e.g., $0.38 < F < 0.4$ for a (14,14) BNNT. Because the larger BNNT under the same applied field is subjected to the larger electric potential than the small one that causes a significant oscillation in $F$-dependent band gaps. Except $F_{1c}^A$, there exists the second critical field inducing a special jump structure in $C_v$-$F$ curves, e.g., $F_{2c}^A = 0.3615$ for a (13,13) BNNT in Fig. 3(b). At the same time the third critical field $F_{3c}^A = 0.3908$ inducing a giant specific heat is also clearly observed for a (14,14) BNNT (Fig. 3(c)). Also notice that the peak height of special jump structures in $C_v$-$F$ curves increases with increasing critical electric field. The main reason is the stronger critical field could flatten edge-states and then induce the larger DOS. That fully reflects the stronger dependence of electronic structures on electric field for larger armchair BNNTs. Thus, larger armchair BNNTs are suitable to create the GESH by applying electric field. Its value is comparable to that induced by the phonon [16-17] and could not be ignored for thermal properties.

As for zigzag BNNTs, the $F$-dependent specific heat of a (21,0) BNNT is first studied. Fig. 3(d) shows that a very small specific heat $C_v = 9.1$ at $T = 1$ $K$, due to the larger band gap (Fig, 2(b)), is induced by the first critical field $F_{1c}^Z = 0.3040$. The features of $C_v$-$F$ curves with increasing temperature are similar to those of a (12,12) BNNT, might be different from the delta-function-like structures for zigzag CNTs induced by linear energy dispersions. The $F$-dependent electronic properties of ZBNNTs, unlike ABNNTs, do not follow a regular rule as the radius increases that is also reflected in $C_v$-$F$ curves. For example, a (22,0) BNNT with $m = 3I + 1$, as mentioned in the above, has a larger band gap leading to a vanishing specific heat at $T < 3\ K$ and $F < 0.32$, as shown in the left inset of Fig. 3(d). The specific heat, even at higher temperature $T = 5\ K$, increasingly reaches to $C_v = 800$ which is one order of magnitude less than those of (21,0) and (23,0) BNNTs. As for a (23,0) BNNT, field-reduced band gap is smaller than that of a (22,0) BNNT such that at $T = 2\ K$ the critical field $F_{1c}^Z = 0.2736$ could induce a specific heat with a value $C_v = 102$. However, except the first critical field, there is no critical field to induce giant specific heat as electric field increases to $F = 0.4$. The second critical field less than $F = 0.4$ appears just for larger BNNTs, e.g., $F_{2c}^Z = 0.3308$ for a (24,0) BNNT shown in the right inset of Fig. 3(d). Therefore, it is much more easily to induce many special structures in specific heat for ABNNTs than ZBNNTs with large radii. The result, for ABNNTs, is beneficial to the modulation of thermal properties by applying electric field.

From the above, the differences of $F$-modulated energy dispersions between a $(m,m)$ ABNNT and a $(m,0)$ ZBNNT are pronouncedly reflected in the $F$-dependent specific heat. They include (I) for the former, the threshold electric field acquired to induce the specific heat is smaller at low temperature ($T < 2\ K$); (II) the magnitude of giant specific heat, for ABNNTs, induced by critical field is reduced with increasing nanotube's radius, but it depends on the modulus of $m$ with respect to three for the later; (III) it is much more easily to induce many special structures in $C_v$-$F$ curves with increasing radius for the former.

On the other hand, the distortion of energy dispersions induced by electric field could also be validated according to the dependence of $C_v$ on temperature. In Fig. 4(a) for a (12,12) BNNT, the specific heats, at non-critical fields, have small values and show nearly linear (at $F = 0.31$) and linear (at $F = 0.32$ and $F = 0.34$) $T$-dependences. Because electric field, increasing from $F = 0.31$ to $F = 0.34$, gradually exhibits linear energy dispersions close to the Fermi energy and reduces the band gap ($\sim 10^{-4}\ eV$), as shown in the inset. While applied field is at the critical value $F_{1c}^A = 0.3079$, much more electronic states get together close to the Fermi energy such that $C_v$ grows quickly at $T > 1\ K$ and shows non-linear $T$-dependence. As for the (21,0) BNNT, Fig. 4(b) shows that the non-vanishing $C_v$ at $F_{1c}^Z = 0.3040$, due to the larger band gap, occurs at higher temperature and has similar dependence on $T$ as compared with that in the (12,12) BNNT. However, at non-critical fields $F = 0.32 - 0.36$ the energy dispersions keep parabolic form and have larger band gaps so that $C_v$-$T$ curves show non-linear behaviors and small values (in the inset). Therefore, the enhancement of critical fields on electronic specific heat is more significant and not to be negligible with increasing temperature.

As nanotube's radius increases, except the first critical field, the second one with larger field strength also enhances specific heat. For (13,13) and (14,14) BNNTs, $C_v$'s, at critical fields $F_{1c}^A$ and $F_{2c}^A$, have similar behaviors with increasing $T$ shown in Fig. 4(c). The two $C_v$-$T$ curves created by $F_{1c}^A$ and $F_{2c}^A$ respectively cross at certain temperature $T_c$. At $T < T_c$ ($T > T_c$), the value of $C_v$ for $F_{1c}^A$ is larger (smaller) than that for $F_{2c}^A$. It is because a modulation of $F_{2c}^A$ on energy dispersion extends to higher energy subbands leading to a more significant enhancement of $C_v$ at high temperature. As for larger ZBNNTs, $C_v$-$T$ curves of (22,0) and (23,0) BNNTs in Fig. 4(d) are non-linear and $C_v$'s have smaller values, since the first critical field exhibits larger band gap and parabolic subbands. It is noticeable that there are big differences in $C_v$-$T$ curves of a (24,0) BNNT created by $F_{1c}^Z$ and $F_{2c}^Z$ respectively. The main reason is that $F_{2c}^Z$ does not exhibit more electronic states close to the Fermi energy. The above clearly shows that $C_v$-$T$ curves have the strong geometry-dependence as well as the field-modulated energy dispersions.

For carbon nanotubes, the magnetic phase, due to their cylindrical symmetry, could induce the quantum interference which strongly affects electronic structures near the Fermi energy; it causes an oscillatory behavior in energy dispersions and band gaps, i.e., the Ahanorov-Bohm oscillations. However, BNNTs have large band gaps and strong ionicity so that effects of magnetic field on electronic structures are less significant. Modulations of magnetic field on electronic structures and thermal properties will be possible if a BNNT is just under critical electric fields. For magnetic flux $\phi = 1$ $\phi_0$, the corresponding magnetic field strength is about 938 $T$ for the (12,12) (or (21,0)) BNNT ($r = 8.3$ Å) that is difficultly achieved in experiments. The smaller magnetic flux $\phi \leq 0.05$ $\phi_0$ ($B \leq 47$ $T$) is considered and the spin-B interaction (Zeeman energy) is included which is defined as $E_z = (g\sigma/m^*r^2)(\phi/\phi_0)$. The g factor is taken to the same that ($\approx 2$) of the pure graphite, $\sigma = \pm\frac{1}{2}$ is the electron spin, and $m^*$ is the bare electron mass. It causes a energy shift with the value $E_z \approx \pm5.5 \times 10^{-3}$ $eV$ at $\phi = 0.05$ $\phi_0$. First we study the change in specific heat of a (12,12) BNNT with increasing magnetic flux at $F_{1c}^A = 0.3079$ and different temperatures. Figure 5(a) shows that the AB-oscillation of electronic properties induced by magnetic phase is also reflected in oscillatory behaviors of $C_v$-$\phi$ curves. With increasing magnetic flux, the value of $C_v$ at $T > 2$ $K$ is smaller than that at $\phi = 0$, whereas it could be enhanced as $T \leq 2$ $K$. Especially, the value of specific heat at $T = 1$ $K$ increases from $C_v = 153$ at $\phi = 0$ to a maximum $C_v = 2256$ at a critical magnetic flux $\phi_c = 0.0086$ $\phi_0$. For magnetic flux $\phi = 0.0086$ $\phi_0$, the corresponding magnetic field strength is $B \approx 8$ $T$ which could be easily performed in experiments. The result reveals that magnetic flux significantly affects electronic states close to the Fermi energy and enhances the low-temperature GESH induced by critical electric field again. The similar result is also found in the zigzag BNNTs, e.g., a (21,0) BNNT in Fig. 5(b). The value of specific heat, at $T = 1$ $K$, increases from $C_v = 0$ at $\phi = 0$ to a maximum $C_v = 2048$ at $\phi_c = 0.0127$ $\phi_0$.

Since the effect of magnetic flux on specific heat is more significant at low temperature, the $\phi$-dependent specific heats at $T = 1$ $K$ are further studied for larger BNNTs at different critical electric fields. For a (13,13) BNNT, magnetic flux could re-enhance the GESHs at the first and second critical fields, as shown in Fig. 5(c). Especially, at $F_{2c}^A = 0.3615$, $C_v = 112$ ($\phi = 0$) reaches to a very large value $C_v = 2888$ at $\phi_c = 0.0142$ $\phi_0$. However, $C_v$-$\phi$ curves of a (14,14) BNNT show different behaviors between critical fields $F_{1c}^A$ and $F_{2c}^A$. At $F_{2c}^A = 0.3277$ the GESH is enhanced with increasing $\phi$ while it is reduced at $F_{1c}^A = 0.2576$. From the above, it is found that an increases of magnetic flux, at the first electric field, could easily enhance the low-temperature GESH for smaller BNNTS, e.g., (12,12) and (13,13) BNNTs. The main reason is Zeeman energy is inversely proportional to the radius squared. The smaller BNNTs through larger Zeeman energy could have much more electronic states close to the Fermi energy. Moreover, Fig. 5(c) also shows that the value of the maximum $C_v$ and the $\phi_c$ is larger at $F_{2c}^A$ than $F_{1c}^A$. The smaller band gap at $F_{2c}^A$ is the main reason.

Different from ABNNTs, magnetic flux always enhances the low-temperature GESH of ZBNNTs while they are under the first electric fields, shown in Fig. 5(d). For a (22,0) BNNT with the widest band gap, the required critical magnetic flux $\phi_c$ inducing a maximum of $C_v$ is the largest while it is the smallest for a (24,0) BNNT with the narrowest band gap. In contrast to a (13,13) BNNT (Fig. 5(c)), the $\phi_c$ of a (24,0) BNNT has a larger value at $F_{2c}^Z$ but the maximum of $C_v$ is smaller as compared with that at $F_{1c}^Z$. It is related to the feature of field-modulated energy dispersion, except band gaps. From the above, the modulation of external fields on thermal properties has strong geometry-dependence that may be used to identify the different geometric structures of BNNTs.

## 4. Conclusion

In this work, the tight-binding model is used to study electronic structures and the low-temperature electronic specific heat of ABNNTs and ZBNNTs in the presence of electric field. Electric field significantly affects energy dispersions, edge-states, and band gaps.

When BNNTs are under a critical electric field, the more electronic states emerge close to the the Fermi energy leading to the large density of states and the giant specific heat. The specific heat has a value comparable to that induced by phonon and reveals strongly non-linear dependence on temperature. The critical field strength, and the value and temperature-dependence of the giant specific heat are profoundly dependent on nanotube's geometry. Additionally, under $F_c$'s, the additional magnetic flux further modulates low-energy dispersions of BNNTs. The interplay between the AB-oscillation and the Zeeman splitting strongly depend on nanotube's geometry. Consequently, the low-temperature specific heats, at $F_c$'s and $\phi = 0$, are always enhanced by certain magnetic flux for ZBNNTs but they are enhanced or diminished for ABNNTs. The modulation of electric and magnetic fields on electronic and low-temperature thermal properties is strongly dependent on the field strength and nanotube's geometry that is expected to be verified by experiments.

**Acknowledgments**

This work was supported in part by the Ministry of Science and Technology of Taiwan, the Republic of China under Grant numbers NSC 101-2112-M-145-001-MY3 and MOST 104-2112-M-145-003.

## Figure Captions

FIG. 1  (a) and (b) are low-energy band structures and the density of states at various electric fields for a (12,12) BNNT. (c)-(d) and (e)-(f) are for (13,13) and (14,14) BNNTs, respectively.

FIG. 2  The same plots as Fig. 1, but for (21,0), (22,0), and (23,0) BNNTs.

FIG. 3  The electric-field-dependent electronic specific heats at different temperatures are shown for (a) (12,12), (b) (13,13), (c) (14,14), and (d) (21,0) BNNTs. The inset in (a) is the detailed $C_v$-$F$ curve of a (12,12) BNNT at $T = 1$ $K$. The $C_v$-$F$ curves at different temperatures of (22,0) and (24,0) BNNTs are shown in the two insets of (d).

FIG. 4  The $T$-dependent electronic specific heats at various electric fields are shown for (a) (12,12) and (b) (21,0) BNNTs. The inset in (a) is the detailed energy dispersion close to the Fermi energy for $F = 0.31$, $F = 0.32$, and $F = 0.34$. The inset in (b) is the detailed $C_v$-$T$ curves at larger electric fields for a (21,0) BNNT. The $T$-dependent electronic specific heats at the first and second critical electric fields for (13,13) and (14,14) BNNTs are shown in (c). (d) is the same plot as (c) but for (22,0), (23,0), and (24,0) BNNTs.

FIG. 5  The $\phi$-dependent electronic specific heats including the spin-B interaction are shown for (a) (12,12) and (b) (21,0) BNNTs at the first critical electric field and various temperatures. The $\phi$-dependent electronic specific heats at the first and second critical electric fields and $T = 1$ $K$ for (13,13) and (14,14) BNNTs are shown in (c). (d) is the same plot as (c) but for (22,0), (23,0), and (24,0) BNNTs.

Fig. 1

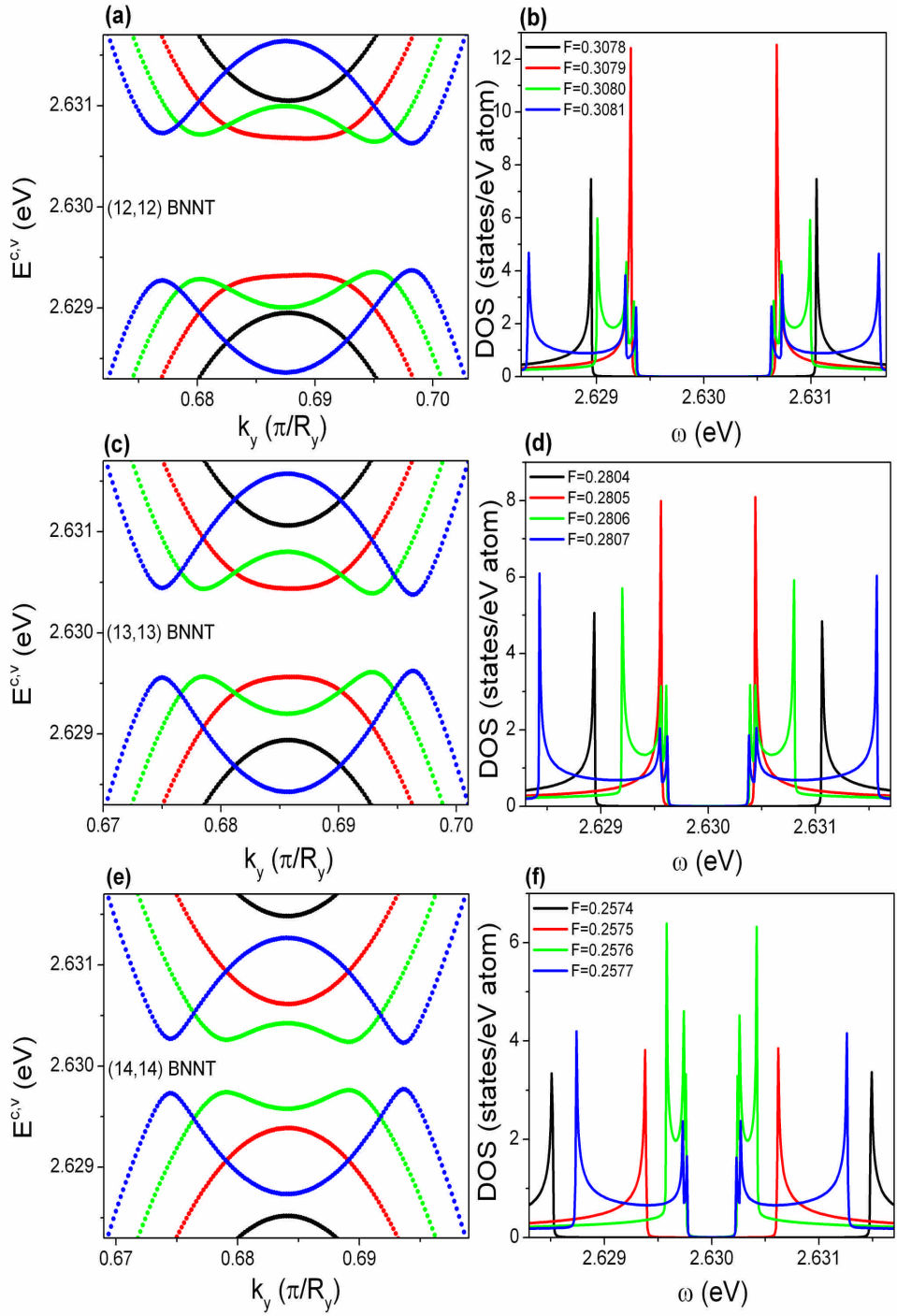

Fig. 2

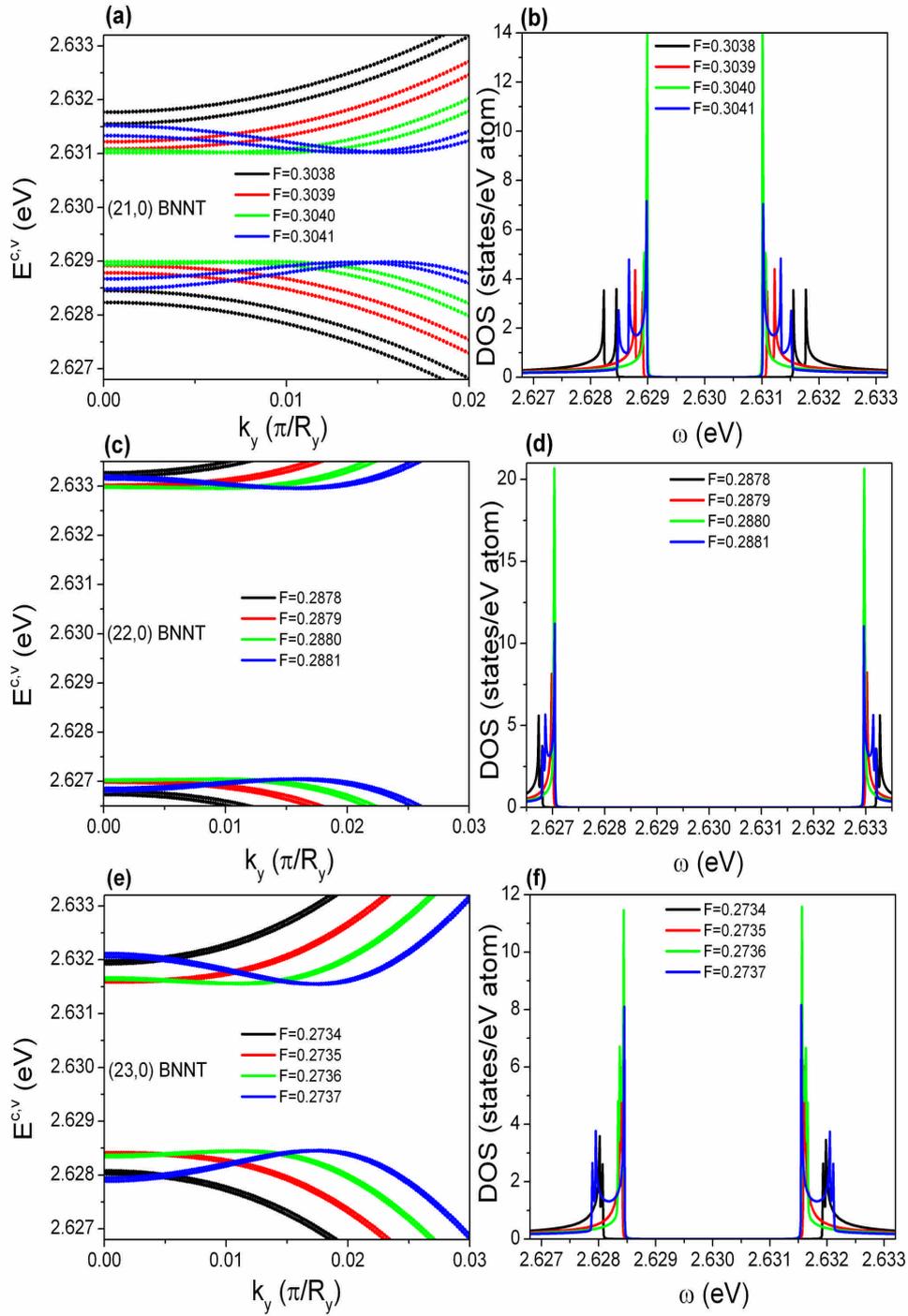

Fig. 3

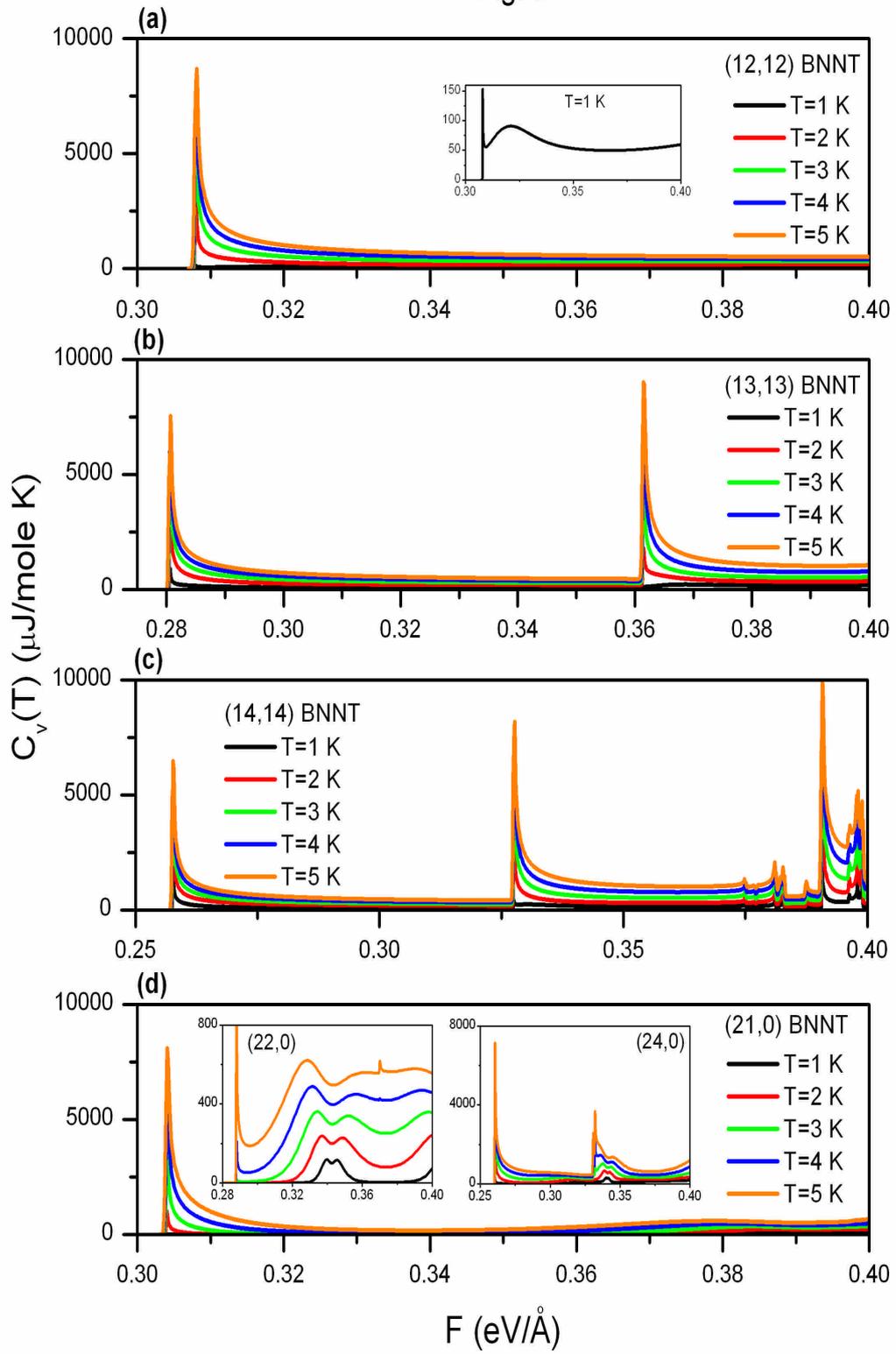

Fig. 4

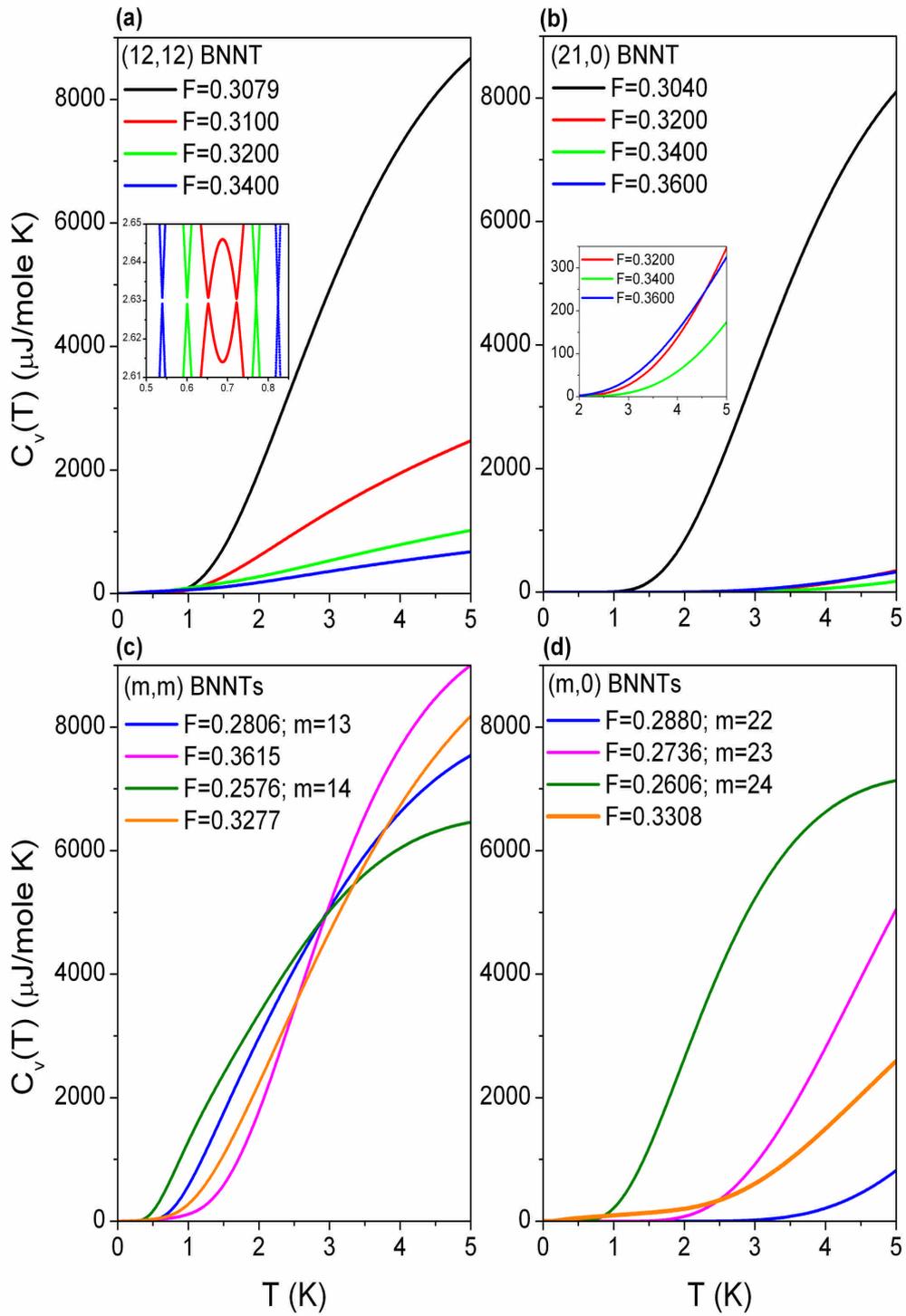

Fig. 5

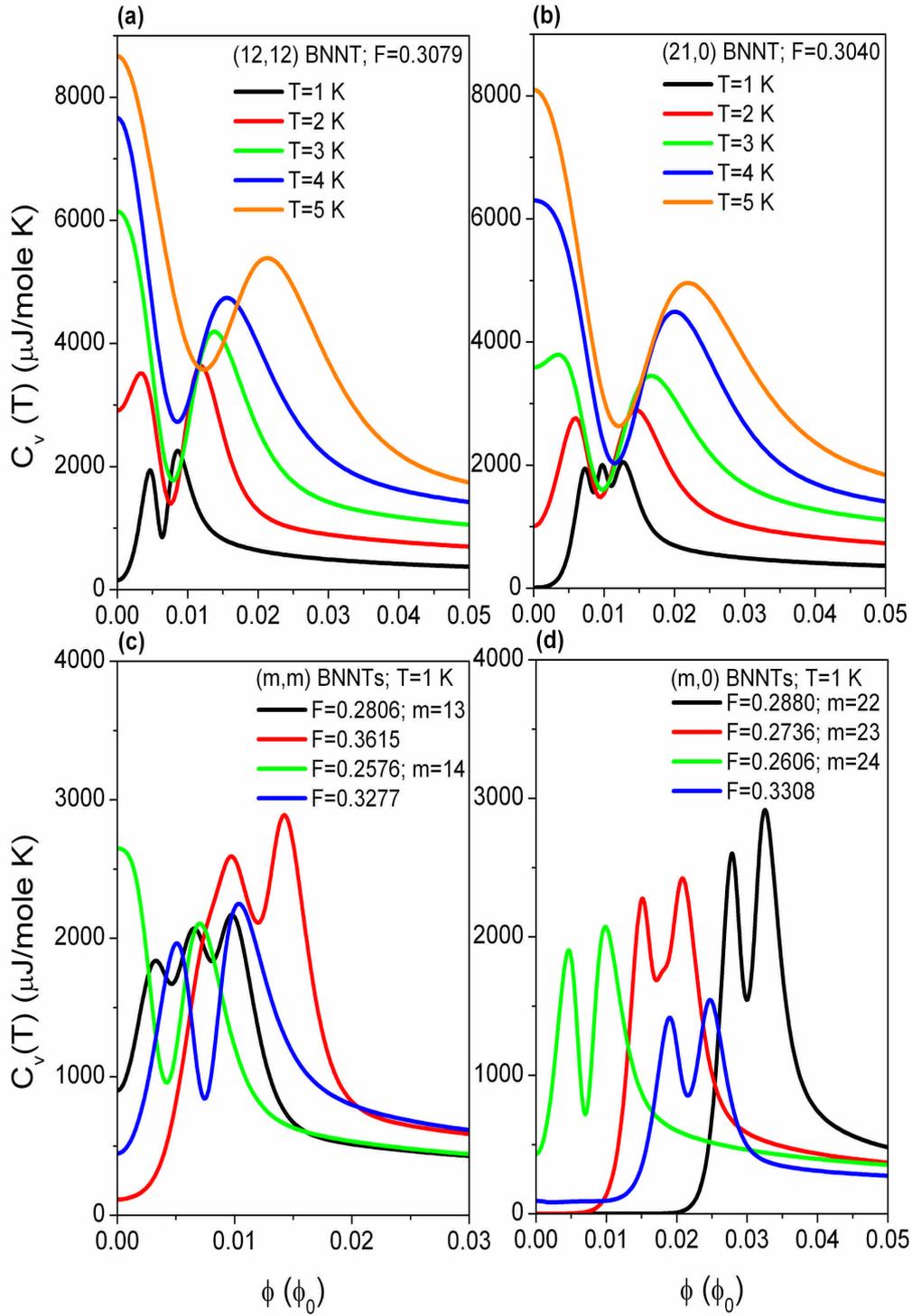